\documentclass[twocolumn,aps,superscriptaddress,showpacs,floatfix,prc]{revtex4}
\usepackage{mathrsfs}
\usepackage{amssymb}
\usepackage{amsmath}
\usepackage{graphicx}
\usepackage[normalem]{ulem}
\usepackage[dvips]{color}
\usepackage{bm}
\usepackage{longtable}

\renewcommand\sout{\bgroup \color{red} \ULdepth=-.5ex \ULset}

\renewcommand{\rm}[1]{\textrm{#1}}

\def\esym{$E_{sym}(\rho)$~}
\def\rpi {$\pi^-/\pi^+$ ratio~}
\def\D {$\Delta(1232)$ resonance~}

\begin{document}
\title{Symmetry potential of $\Delta(1232)$ resonance and its effects on the $\pi^-/\pi^+$ ratio
in heavy-ion collisions near the pion production threshold}

\author{Bao-An Li\footnote{Email: Bao-An.Li@Tamuc.edu}}
\affiliation{Department of Physics and Astronomy, Texas A$\&$M University-Commerce, Commerce, TX 75429-3011, USA}

\begin{abstract}
Effects of the completely unknown symmetry (isovector) potential of the \D on the total and differential \rpi in heavy-ion collisions at beam energies from 100 to 1000 MeV/A are explored within an isospin-dependent transport model IBUU. The effects are found to be negligible at beam energies above the pion production threshold due to the very short lifetimes of less than 2 fm/c for $\Delta$ resonances with masses around $m_{\Delta}=1232$ MeV, leaving the $\pi^-/\pi^+$ ratios of especially the energetic pions still a reliable probe of the high-density behavior of nuclear symmetry energy
$E_{sym}(\rho)$. However, as the beam energy becomes deeply sub-threshold for pion production, effects of the $\Delta$ symmetry potential becomes appreciable especially on the \rpi of low-energy pions from the decays of low-mass $\Delta$ resonances which have lived long enough to be affected by their mean-field potentials, providing a useful tool to study the symmetry potential and spectroscopy of $\Delta$ resonances in neutron-rich nuclear matter.  Interestingly though, even at the deeply sub-threshold beam energies, the differential \rpi of energetic pions remains sensitive to the \esym at supra saturation densities with little influence from the uncertain symmetry potential of the $\Delta$ resonance.
\end{abstract}

\maketitle

\section{Introduction}
Thanks to the great efforts of many people in both nuclear physics and astrophysics communities, much progress has been made in recent years in constraining the density dependence of nuclear symmetry energy \esym around the saturation density $\rho_0$, see, e.g., refs. \cite{lkb98,liudo,Steiner05,Bar05,LCK,Trau12,Tsang12,Jim13,LiBA13,Hor14,EPJA}  for comprehensive  reviews. At supra saturation densities, however, even the tendency of the \esym is still controversial based on the model analyses of very limited data available. Moreover, predictions on the high-density behavior of nuclear symmetry energy based on both microscopic many-body theories and phenomenological models diverge broadly mainly because of our poor knowledge about the spin-isospin dependence of nuclear many-body forces, isospin-dependence of short-range correlations, the nature of tensor forces, etc, besides the well-known difficulties of treating precisely quantum many-body problems \cite{EPJA}.

Heavy-ion collisions are the only means in terrestrial laboratories to create dense nuclear matter. To pin down the high-density symmetry energy has been one of the major goals of the low-intermediate energy heavy-ion reaction community in recent years. Among the potential probes of the high-density symmetry energy, the \rpi in heavy-ion collisions has attracted much attention since it was first proposed \cite{Li02}. Unfortunately,  several model analyses \cite{Xiao09,Feng10,Xie13,Rus11,Coz13,Cozma14} of some existing data have so far been inconclusive. Realizing the importance of thoroughly understanding all aspects of the $\pi-N-\Delta$ dynamics in isospin-asymmetric matter especially since several experiments measuring the \rpi are currently underway at RIKEN \cite{Tsang} and NIRS \cite{Sako14} besides those planned at other facilities \cite{Xiao14,Hong14}, significant efforts have been made recently in studying effects of the in-medium pion production threshold \cite{Fer05}, pion mean-field \cite{Xu10,Xu13,Tae14,Hon14}, electromagnetic field \cite{Ouli11}, neutron-skin \cite{Wei14} and nucleon-nucleon short-range correlation \cite{Li14,Yong15} on pion production mostly at intermediate energies above the pion production threshold of about 300 MeV/A. Extending these efforts, we report here results of a study on effects  of the completely unknown symmetry (isovector) potential of the \D on both the total and differential $\pi^-/\pi^+$ ratios in heavy-ion collisions at beam energies from 100 to 1000 MeV/A within an isospin-dependent transport model IBUU \cite{Li02}. It is found that the $\pi^-/\pi^+$ ratio of energetic pions remain a useful probe of the high-density \esym without much influence by the completely unknown symmetry potential of the $\Delta$ resonance.

\section{Modeling the symmetry potentials of nucleons and $\Delta(1232)$ resonances}
\begin{figure}[htb]
\begin{center}
\includegraphics[width=9cm,height=8cm]{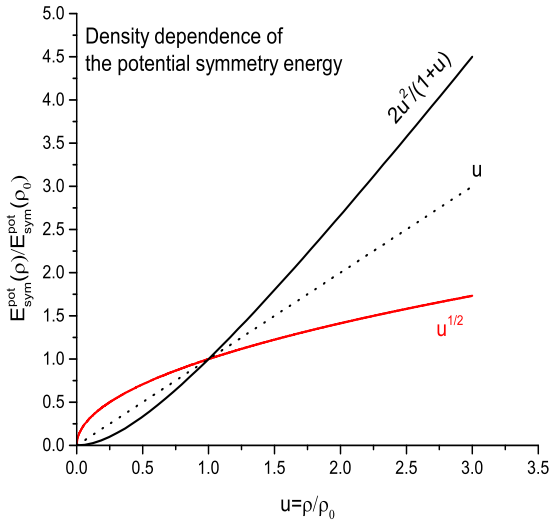}
\caption{(Color online) \label{Esym} The density dependence of the potential symmetry energy used in this study.}
\end{center}
\end{figure}
We first recall here briefly the major ingredients and new modifications of the IBUU transport model that are relevant for this study. First of all, to ease future comparisons with studies using other transport models, for the isoscalar potential we use the momentum-independent BKD (Bertsch-Kruse-Das Gupta) nucleon potentials \cite{Ber84} with an incompressibility of $k_0=236$ MeV that is a common option in most transport models developed for investigating heavy-ion reactions. This choice does not affect our qualitative conclusions as we are dealing with the ratios of charged pions that are insensitive to the nucleon isoscalar potential. On the other hand, it allows us to generate timely enough reaction events necessary for studying various cases of deeply sub-threshold pion production using the best high-performance computing resources available to the author today.
We caution, however, to compare with data of ongoing experiments to constrain quantitatively the \esym at supra saturation densities using the differential \rpi of energetic pions, the momentum dependence of both the isoscalar and isovector potentials of nucleons and $\Delta$ resonances should be considered carefully.

For the purpose of this work, it is sufficient and more economic to use the simplest parameterization for the density dependence of nuclear symmetry energy $E_{sym}(\rho)$. Since the earlier studies about effects of nuclear symmetry energy on several properties of neutron stars \cite{Pra88b,Lat91} and the isospin dynamics of heavy-ion collisions \cite{LiBA97a}, one often parameterizes the \esym as
\begin{eqnarray}\label{esymfu}
E_{\rm sym}(\rho)=(2^{2/3}-1)\frac{3}{5}E_{F}^{0}[u^{2/3}-F(u)]
+E_{\rm sym}(\rho_{0})F(u),
\end{eqnarray}
with the typical function $F(u)$ being
\begin{eqnarray}\label{fu}
F_1(u)=\frac{2u^2}{1+u},~F_2(u)=u,~F_3(u)=u^{1/2},
\end{eqnarray}
where $u\equiv \rho/\rho_0$ is the reduced baryon density, $E_F^0=36$ MeV is the Fermi energy and $E_{\rm sym}(\rho_0)=31$ MeV is the symmetry energy at $\rho_0$. The selected three forms of the
F(u) span a large uncertain range of $E_{sym}(\rho)$. What is important for the present study is the potential symmetry energy given in the second part of Eq. \ref{esymfu} and shown in Fig. \ref{Esym}.  Assuming there is no momentum dependence, the nucleon symmetry potential $v_{\rm asy}(n/p)$ up to the linear term in the isospin asymmetry $\delta\equiv (\rho_n-\rho_p)/(\rho_n+\rho_p)$ of nucleonic matter can be obtained from taking the partial derivative of the potential symmetry energy density with respect to the neutron or proton density \cite{LCK}. The $v_{\rm asy}(n/p)$ corresponding to the three functions of $F(u)$ is  given by, respectively,
\begin{eqnarray}\label{vasy}
v_{\rm asy1}(n/p)&=&\pm 4e_a \frac{u^2}{1+u}\delta,\nonumber\\
v_{\rm asy2}(n/p)&=&\pm 2 e_a u\delta,\nonumber\\
v_{\rm asy3}(n/p)&=&\pm 2e_a
u^{1/2}\delta
\end{eqnarray}
where the $\pm$ sign is for neutrons/protons and $e_a\equiv E_{\rm sym}(\rho_0)-(2^{2/3}-1)\,{\textstyle\frac{3}{5}}E_F^0$. We notice here that the potential symmetry energy density for $F_1(u)$ and $F_3(u)$ also contributes a term proportional to $\delta^2$ to the isoscarlar potential. However, such a term in heavy-ion collisions is negligibly small compared to the isoscalar potential corresponding to the EOS of symmetric matter.

Associated with the $\pi$-N-$\Delta$ dynamics in nuclei, nuclear reactions and neutron stars, the isoscalar potential for the \D has been extensively studied for a long time using various many-body theories and interactions, see, e.g., refs. \cite{Brown75,Oset81,Tak85,Oset87, Con90, Bal90,Jon92,Baldo94}. For a comprehensive review of the relevant issues, we refer the reader to ref. \cite{pion-book}. Indeed, there are indications from analyses of electron-nucleus, photoabsorption and pion-nucleus scattering that the $\Delta$ isoscalar potential $v_{\Delta}$ is in the range
of $-30\,\mathrm{MeV}+v_{\textrm{N}}\le v_{\Delta}\le v_{\textrm{N}}$ with respect to the nucleon isoscalar potential
$v_{\textrm{N}}$\,\cite{Drago14}.  Since a constant potential has no dynamical effect in heavy-ion reactions and there is no information available as to how the difference in the isoscalar potentials for nucleons and $\Delta$ resonances depend on the density and/or other properties of the medium, in this study we thus use the same isoscalar potentials for nucleons and $\Delta$ resonances. The mean-field potential of pions in neutron-rich matter 
is an interesting issue. While there is some information about the strength of pion potential in nuclei from study pionic atoms, there is little information about its density, isospin and momentum dependence in nuclear matter.
Recognizing its possible importance but also considering the currently existing controversies about it, we neglect here the pion mean-field.

As pointed out already in ref. \cite{Cozma14,Drago14,Cai14}, we essentially know nothing about the isovector (symmetry) potential of the \D in isospin-asymmetric nuclear matter. In heavy-ion collisions near the pion production threshold, the $\Delta$ population is very small. For example, for $^{132}$Sn+$^{124}$Sn reaction at a beam energy of 400 MeV/nucleon and an impact parameter of 1 fm,  the maximum total multiplicity of $\Delta$ resonances in all charge states reached during the reaction is about 4 \cite{LYZ}. This is about 3.2\% of the total number of nucleons involved. Thus, the mean-field potential of $\Delta$ is mainly due to its interactions with nucleons not other $\Delta$ resonances. Therefore, similar to nucleons, the isovector potential of $\Delta$ resonances is also proportional to the isospin asymmetry $\delta$ of nucleons due to the isospin-dependent $\tau_3(\Delta)\cdot \tau_3(N)$ term in the $\Delta-N$ interaction \cite{Lenske}. The population of $\Delta$ only has a small perturbative effect on the  isospin asymmetry $\delta$ of nucleonic matter. Assuming the $\rho$ meson exchange among nucleons and baryon resonances is mainly responsible for the isovector interactions, there is a simple scaling relation between the $NN\rho$ and $\Delta\Delta\rho$ coupling constants $g_{\rho\textrm{N}}$ and $g_{\rho\Delta}$ in free-space based on the quark model of baryons \cite{Brown75}. However, the relationship between the isovector potentials of nucleons and baryon resonances in isospin-asymmetric matter is completely unknown. In fact, the critical density for \D formation in neutron stars has been found to depend almost linearly on the unknown ratio $x_{\rho}\equiv g_{\rho\Delta}/g_{\rho\textrm{N}}$ \cite{Drago14,Cai14}. Depending on the assumed value of $x_{\rho}$, the formation of \D can even happen at $\rho_0$ and affect significantly properties of neutron stars \cite{Cai14}. In addition, the isovecor Lorentz-scalar $\delta$-meson may also play some roles in determining the baryon isovector potentials. However, its couplings with baryons are even less known. Thus, because of their ramifications in both nuclear physics and astrophysics both the isoscalar and isovector parts of the $\Delta$ potential deserve further investigations. In particular, it would be interesting to know if any observable in heavy-ion collisions especially those induced by neutron-rich nuclei may be used to probe the $\Delta$ isovector potential in neutron-rich matter.

In two earlier studies \cite{Li02,Mah98}, the isovector potential of $\Delta$ resonances was connected to that of nucleons based on rather different assumptions. In ref. \cite{Li02} and subsequently many other studies, considering the $\Delta$ as a molecule consisting of a nucleon and a pion, the isovector potential of the $\Delta$ resonance is an average of that for neutrons and protons with weights given
by the square of the Clebsch-Gordon coefficients in the $\Delta\leftrightarrow \pi N$ processes conserving the total isospin. Under these assumptions, one has \cite{Li02}
\begin{eqnarray}\label{dpot1}
v_{asy}(\Delta^-)&=&v_{asy}(n),\nonumber\\
v_{asy}(\Delta^0)&=&\frac{2}{3}v_{asy}(n)+\frac{1}{3}v_{asy}(p)=\frac{1}{3}v_{asy}(n),\nonumber\\
v_{asy}(\Delta^+)&=&\frac{1}{3}v_{asy}(n)+\frac{2}{3}v_{asy}(p)=-\frac{1}{3}v_{asy}(n),\nonumber\\
v_{asy}(\Delta^{++})&=&v_{asy}(p)=-v_{asy}(n).
\end{eqnarray}
Assuming that baryon isovector potentials are all due to the $\rho$ meson exchange neglecting possible contributions of the $\delta$-meson, the above prescription is equivalent to setting $x_{\rho}=1/3$. In line with the above assumption, taking into account the small contributions of the $\Delta$ resonances the effective isospin asymmetry $\delta_{like}$ of the excited baryonic matter can be obtained from
\begin{equation}
\delta_{like}\equiv \frac{(\rho_n)_{like}-(\rho_p)_{like}}
{(\rho_n)_{like}+(\rho_p)_{like}}
\end{equation}
where
\begin{eqnarray}
(\rho_n)_{like}&=&\rho_n+\frac{2}{3}\rho_{\Delta^0}+\frac{1}{3}\rho_{\Delta^+}
+\rho_{\Delta^-},\\\
(\rho_p)_{like}&=&\rho_p+\frac{2}{3}\rho_{\Delta^+}+\frac{1}{3}\rho_{\Delta^0}
+\rho_{\Delta^{++}}.
\end{eqnarray}
It is evident that the $\delta_{like}$ reduces naturally to the isospin asymmetry $\delta$ of nucleonic matter as the
beam energy becomes smaller than the pion production threshold. 

While in ref. \cite{Mah98}, the following prescription according to the effective charge of the \D was used
\begin{eqnarray}\label{dpot2}
v_{asy}(\Delta^-)&=&2v_{asy}(n)-v_{asy}(p)=3v_{asy}(n),\nonumber\\
v_{asy}(\Delta^0)&=&v_{asy}(n),\nonumber\\
v_{asy}(\Delta^+)&=&v_{asy}(p)=-v_{asy}(n),\nonumber\\
v_{asy}(\Delta^{++})&=&2v_{asy}(p)-v_{asy}(n)=-3v_{asy}(n).
\end{eqnarray}
This prescription is equivalent to setting $x_{\rho}=1$, thus the magnitudes of these isovector potentials for \D are 3 times that of Eq. \ref{dpot1}. We emphasize that the ratios of $\Delta$ isovector potentials for different charge states are the same in both prescriptions.  In this work, we use the symmetry potentials in Eq. \ref{dpot1} as a basis and multiply them with a ``$\Delta$-probing factor $f_{\Delta}$". By varying the value of $f_{\Delta}$ we can explore effects of the isovector potential of \D on pion observables in heavy-ion collisions.

 \begin{figure}[htb]
\begin{center}
\includegraphics[width=13cm,height=10cm]{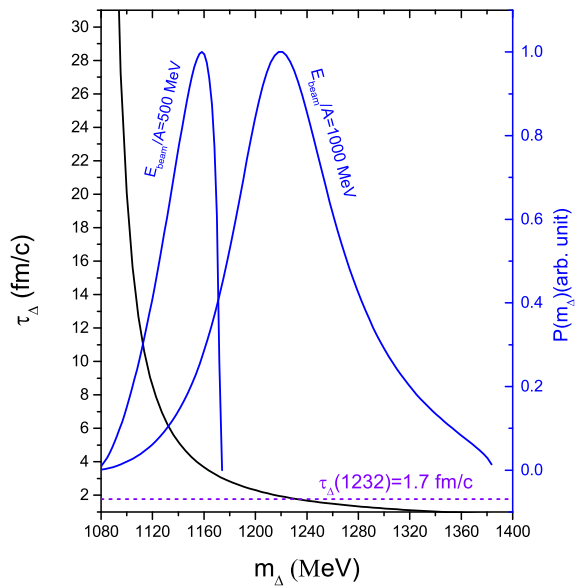}
\caption{(Color online) \label{tau} The mean lifetime (black line) of \D as a function of its mass. The dashed line at 1.7 fm/c indicates the lifetime of the $\Delta$ as its centroid mass of 1232 MeV. The blue lines are the $\Delta$ mass distributions in $NN\rightarrow N\Delta$ collisions at nucleon-nucleon (NN) center of mass energy of $\sqrt s=2.11$ GeV and 2.32 GeV corresponding to first chance NN collisions in nucleus-nucleus reactions at a beam energy of 500 and 1000 MeV/nucleon, respectively.}
\end{center}
\end{figure}

\section{Modeling the mass distribution and lifetime of $\Delta(1232)$ resonances}
Whether the $\Delta$ potential has any observable effect in heavy-ion collisions depends strongly on the lifetime of \D over which the $\Delta$ potential acts to modify its momentum. It is thus important to outline how we
model the mass distribution and the corresponding width of the \D in heavy-ion collisions. As discussed in detail in ref. \cite{Li93} and reviewed in ref. \cite{art}, in the IBUU transport model the mass $m_r$ of a baryon resonance $r$ ($\Delta$ or $N^*$) produced in an inelastic nucleon-nucleon collision $(N+N\rightarrow N^{\prime}+r)$ is generated according to a modified Breit-Wigner function \cite{Dan91}
\begin{equation}
P(m_{r})=\frac{p_f\cdot m_{r}\cdot \Gamma(m_{r})}{(m_{r}^{2}
-m_{r0}^{2})^{2}+m_{r0}^{2}\Gamma_{r}^{2}(m_{r})}
\end{equation}
where $p_f$ is the final nucleon momentum in the NN center of mass frame, $m_{r0}$ and $\Gamma(m_{r})$  are the centroid and width of the resonance, respectively.
As noticed earlier \cite{Dan91}, the $p_f$ surpreses significantly the $\Delta$ production near its maximum mass of $m_{\Delta}(\rm max)=\sqrt s-m_N$ compared to the simple
Breit-Wigner function widely used in many transport models. This effect is especially significant when the $m_{\Delta}(\rm max)$ is less than 1232 MeV in sub-threshold pion production.
For $r=\Delta(1232)$, the width $\Gamma(m_{r})$ is given by \cite{kit}
\begin{equation}
\Gamma(\Delta(1232)) =0.47 q^{3}/ \left(m_{\pi}^{2}+0.6q^{2}\right)~{\rm (GeV)}
\end{equation}
where $q$ is the pion momentum in the $\Delta$ rest frame in the $\Delta\rightarrow \pi+N$ decay process.
During each time step of very small length $dt$, a Monte Carlo sampling of the $\Delta$ decay is carried out according to the probability
\begin{equation}
P_{\rm decay}=1-\exp(-dt/\tau_{\Delta})\approx dt/\tau_{\Delta}
\end{equation}
where $\tau_{\Delta}=\hbar /\Gamma(m_{\Delta})$ is the $\Delta$ mean lifetime. 

It is necessary to mention that there are many interesting issues regarding the $\Delta$ spectroscopy in neutron-rich matter to be studied \cite{Lenske},
such as the possible modification of its width (lifetime) and centroid mass. Moreover, there are already some interesting work on the delay times of $NN\rightarrow N\Delta$ scattering, how the width of \D in the scattering should be modeled quantum mechanically and whether pions should be directly produced without going through the intermediate \D at all in heavy-ion collisions \cite{Dan96,Leu01,Lar02}. These issues are not addressed here. Our calculation of the lifetime of the \D is rather standard used in most transport models. Moreover, as discussed in detail in refs. \cite{Li1,Li2}, in IBUU we use free-space $NN\rightarrow N\Delta$ cross sections and the scattering kinematics is that in vacuum. This is a good approximation when nucleons and $\Delta$ resonances have the same momentum-independent potential, since in this case the baryon potentials before and after the NN collision cancel out in the energy conservation equation. In other cases, however, our treatment is equivalent to neglecting residual effects of the baryon mean-field potential on the NN center of mass energy $\sqrt s$, except those through the momenta of the colliding nucleons from their earlier propagation in the medium. As we noticed earlier, how the isospin and momentum dependent potentials of nucleons and $\Delta$ resonances may affect particle production thresholds is an interesting issue under investigation by several other groups. We also plan to address this issue within the IBUU approach in the near future.
\begin{figure}
\includegraphics[width=10cm,height=15cm]{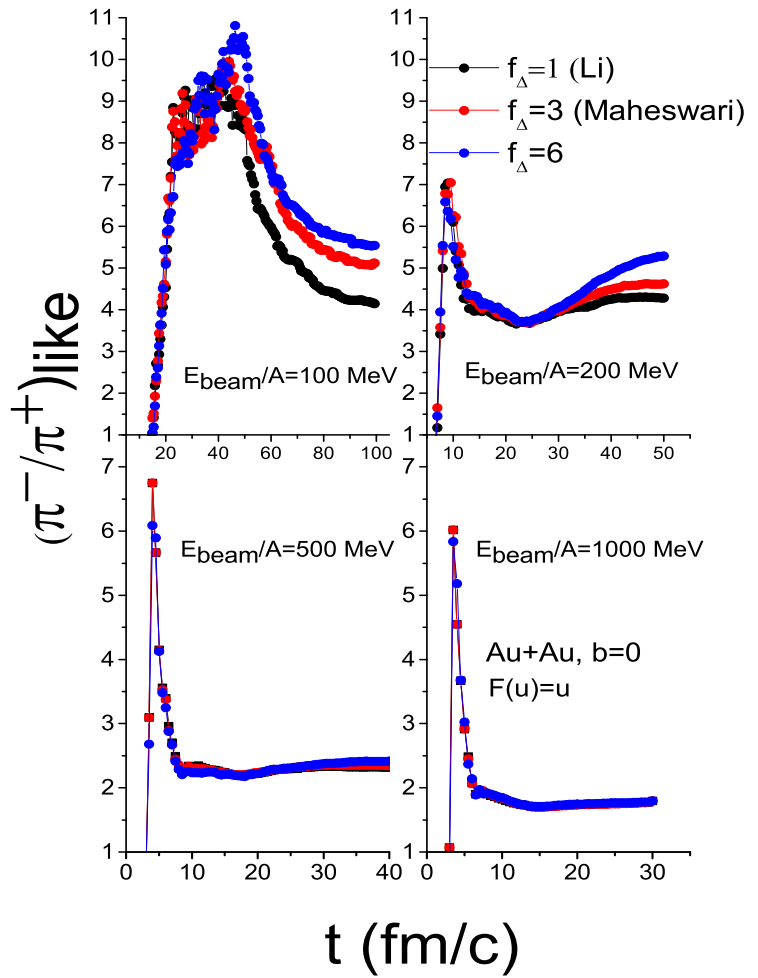}
\caption{(Color online) \label{rpievo} The $(\pi^-/\pi^+)_{like}$ ratio as a function of time in head-on Au+Au reactions at a beam energy of 100, 200, 500 and 1000 MeV/nucleon
with the same F(u)=u but 3 different $f_{\Delta}$ values.}
\end{figure}

\begin{figure}
\begin{center}
\includegraphics[width=10cm,height=9cm]{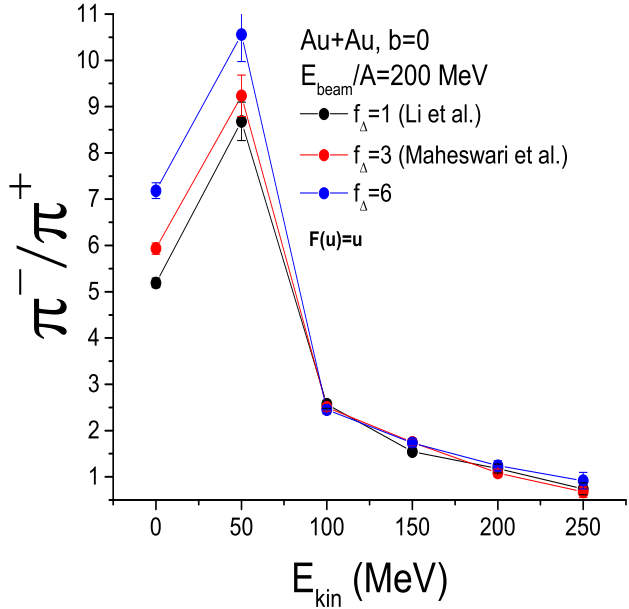}
\caption{(Color online) \label{rpi200} The \rpi as a function of pion kinetic energy in head-on Au+Au reactions at a beam energy of 200 MeV/nucleon with the same F(u)=u but three different isovector potentials for the $\Delta$ resonance.}
\end{center}
\end{figure}

As discussed earlier \cite{bass} and seen clearly in Fig.\ \ref{tau},
the low-mass $\Delta$ resonances predominantly produced in NN collisions near the pion production threshold have very long lifetimes during which the $\Delta$ mean-field may
act. Pions from decays of these resonances may then carry some information about the $\Delta$ potential. On the contrary, as indicated by the dashed line in Fig.\ \ref{tau},
the $\Delta$ resonance around its mass centroid of 1232 MeV has a lifetime of only about 1.7 fm/c which is too short for the $\Delta$ mean-field to cause any significant change in its momentum. Moreover, subsequent $\pi+N\leftrightarrow \Delta$ reaction cycles tend to destroy the memory of the pion about the dynamical history of its parent $\Delta$. The two blue lines are the $\Delta$ mass distributions in $NN\rightarrow N\Delta$ collisions at the NN center of mass energy of $\sqrt s=2.11$ GeV and 2.32 GeV corresponding to the first-chance NN collisions in nucleus-nucleus reactions at a beam energy of 500 and 1000 MeV/nucleon without considering contributions of the initial nucleon Fermi momentum, respectively. It is seen that at 1000 MeV/nucleon, the $\Delta$ mass distribution peaks around 1232 MeV. The $\Delta$ resonances created at this beam energy are thus mostly very short-lived. As the beam energy decreases towards sub-threshold energies, e.g., at 500 MV/nucleon, more long-lived low-mass $\Delta$ resonances are produced. One thus expects that pions from heavy-ion collisions at sub-threshold beam energies to be more sensitive to the $\Delta$ potential than reactions at higher beam energies. Moreover, only a small fraction of nucleons go through the $\Delta$ resonances for a relatively short time interval compared to the whole reaction period over which the nucleon potential acts continuously. We thus do not expect to see any effect of the $\Delta$ potential on nucleon observables.

\begin{figure}
\begin{center}
\includegraphics[width=10.cm,height=9cm]{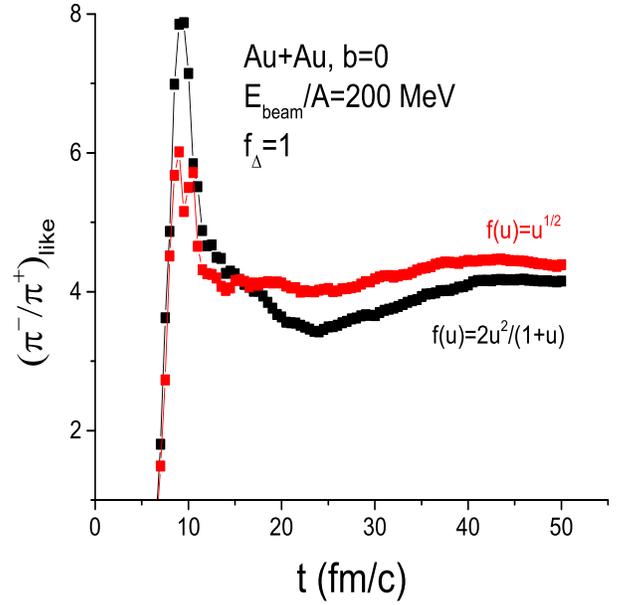}
\caption{(Color online) \label{rpiesym} The $(\pi^-/\pi^+)_{like}$ ratio as a function of time in head-on Au+Au reactions at a beam energy of 200 MeV/nucleon
with $f_{\Delta}=1$ but two different forms for the F(u). }
\end{center}
\end{figure}

\begin{figure}
\begin{center}
\includegraphics[width=9.cm,height=8cm]{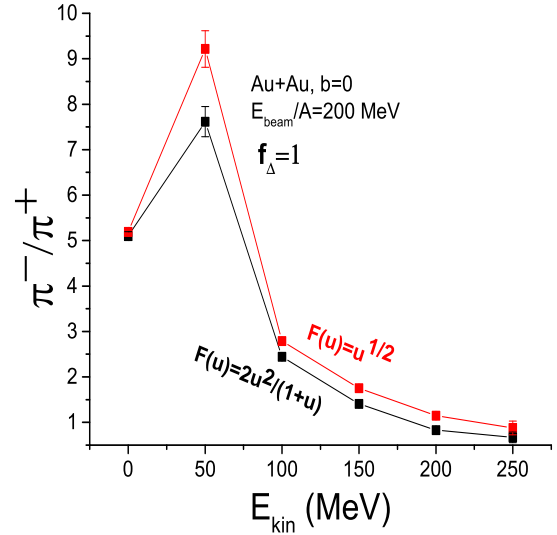}
\caption{(Color online) \label{rpi-k} The $(\pi^-/\pi^+)_{like}$ ratio as a function of pion kinetic energy in head-on Au+Au reactions at a beam energy of 200 MeV/nucleon
with $f_{\Delta}=1$ but two different forms for the F(u). }
\end{center}
\end{figure}

\section{Effects of the symmetry potentials of $\Delta(1232)$ and nucleons on the total and differential $\pi^-/\pi^+$ ratios}
We first examine in Fig.\  \ref{rpievo} the evolution of the $(\pi^-/\pi^+)_{like}$ ratio
\begin{equation}
(\pi^-/\pi^+)_{like}\equiv \frac{\pi^-+\Delta^-+\frac{1}{3}\Delta^0}
{\pi^++\Delta^{++}+\frac{1}{3}\Delta^+}
\end{equation}
for head-on Au+Au reactions at beam energies of 100,  200, 500 and 1000 MeV/nucleon with $F(u)=u$ and $f_{\Delta}=1$, 3 and 6, respectively. Approximately 120,0000 events were generated in each case. The $(\pi^-/\pi^+)_{like}$ ratio naturally becomes the final $\pi^-/\pi^+$ ratio at the freeze-out when the reaction time $t$ is much longer than the lifetime of the $\Delta$ resonance $\tau_{\Delta}$.  The extremely long-lived and very-light $\Delta$ resonances with $m_{\Delta}$ near $m_N+m_{\pi}$ are forced to decay at the freeze-out time if they were not reabsorbed by the $N\Delta\rightarrow NN$ reactions earlier. 
With $f_{\Delta}=1$ and 3, the symmetry potential for the \D is that of Eq. \ref{dpot1} and Eq. \ref{dpot2}, respectively. We also use a large value of $f_{\Delta}=6$ to explore its effects. It is seen that effects of the $\Delta$ symmetry potential only become appreciable when the beam energy approaches the pion production threshold. At deeply sub-threshold beam energies, the effect of the $\Delta$ symmetry potential on the total \rpi is significant.  At $E_{beam}/A=200$ MeV, for example, the total \rpi increases by about 25\% when the $f_{\Delta}$ is increased from 1 to 6.
As one expects, the final $(\pi^-/\pi^+)_{like}$ ratio is higher with a larger $f_{\Delta}$ value which amplifies the difference in potentials for $\Delta^-$ and $\Delta^{++}$. The beam energy dependence of the effect of the $\Delta$ symmetry potential is consistent with our expectations based on the $\Delta$ lifetime shown in Fig.\ \ref{tau} and discussed earlier. Namely, only pions from the decays of light but long-lived  $\Delta$ resonances created in low-energy collisions are affected by the $\Delta$ potential.  Also, consistent with the known systematics of pion production \cite{FOPI}, the \rpi increases with decreasing beam energy.

The results shown in Fig.\ \ref{rpievo} seem to make the current efforts of using the sub-threshold \rpi to constrain the \esym around $1-2\rho_0$ even more challenging unless one knows the exact relationship between the isovector potentials of nucleons and $\Delta$ resonances. To assess the situation more clearly, we examine in Fig.\ \ref{rpi200} the final \rpi as a function of pion kinetic energy for the head-on Au+Au reactions at $E_{beam}/A=200$ MeV. It is interesting to see that effects of the $\Delta$ symmetry potential from varying the $f_{\Delta}$ parameter is mainly on low energy pions below about 70 Mev. The more energetic pions are not much affected at all by the $\Delta$ potential even at such a deeply sub-threshold beam energy. Again, this is due to the mass-dependent lifetime of the $\Delta$ resonance. The low (high) energy pions are mainly from the long- (short-) lived, low (high) mass $\Delta$ resonances. Thus, to investigate the isovector potential of \D one should focus on the \rpi of low energy pions for which the Coulomb fields are also important but are well understood. 

The observation that the uncertain isovector potential of the \D has almost no effect on the differential \rpi of energetic pions is very encouraging. It is thus still possible to use these energetic pions to probe the \esym at supra saturation densities. To be more quantitative and make a comparison, shown in Fig.\ \ref{rpiesym} are the $(\pi^-/\pi^+)_{like}$ ratio as a function of time in the head-on Au+Au reactions at a beam energy of 200 MeV/nucleon with $f_{\Delta}=1$ but two different forms for the F(u). As known before, the total \rpi at the freeze-out has an appreciable sensitive to the variation of F(u). We notice that the current uncertainty of the potential symmetry energy at supra saturation densities going from large negative to positive values is much larger than the difference between the $F_1(u)=2u^2/(1+u)$ and $F_3(u)=u^{1/2}$ used as examples here \cite{LCK}.
More interestingly, as shown in Fig.\ \ref{rpi-k} the differential \rpi of energetic pions still shows an appreciable sensitivity to the variation of F(u). Since these energetic pions are not much influenced by the unkonwn $\Delta$ potential, the observations here similar to the findings from studying the differential \rpi in Sn+Sn reactions at similar beam energies within the PBUU model \cite{Hon14}, indicate that the \rpi of energetic pions is still a useful probe of the \esym at supra saturation densities. 

\section{Summary}
In summary, within the IBUU transport model we studied effects of the completely unknown $\Delta$ symmetry potential on the total and differential \rpi in heavy-ion collisions at beam energies from 100 to 1000 MeV/A. We found that the $\Delta$ symmetry potential has negligible (significant) effect on the integral ratio of charged pion multiplicities at beam energies above (below) the pion production threshold.  In sub-threshold reactions, the differential \rpi at low kinetic energies is also significantly affected by the $\Delta$ symmetry potential, providing a useful tool to study the symmetry potential and spectroscopy of $\Delta$ resonances in neutron-rich nuclear matter. On the other hand, the differential \rpi of energetic pions is little affected by the $\Delta$ symmetry potential even at the deeply sub-threshold beam energies, leaving it still one of the best known probes of the \esym at supra saturation densities. We also found that the mass-dependent $\Delta$ width (lifetime), i.e., the $\Delta$ spectroscopy in neutron-rich matter,  is very important in determining whether the $\Delta$ potential plays any role in heavy-ion collisions.

\section{Acknowledgement}
This work was started to answer several questions raised by Mike Lisa at two occasions first in Nov. of 2009 in Ohio, then in May of 2014 in Catania regarding the dynamical treatment of pions and $\Delta$ resonances in transport models. I would thus like to thank Mike Lisa for raising the important questions, his interest in knowing the answers and the helpful discussions. I would also like to thank P. Danielewicz, B.J. Cai, M.D. Cozma, F.J. Fattoyev and W.G. Newton for helpful discussions. This work was supported in part by the U.S. National Science Foundation under Grant No. PHY-1068022, the U.S. Department of Energy's Office of Science under Award Number DE-SC0013702 and the National Natural Science Foundation of China under grant no. 11320101004.

\end{document}